\documentclass[a4paper,12pt]{article}
\pdfoutput=1

\usepackage{amsmath}
\usepackage{amssymb}
\usepackage[numbers, square, comma, sort&compress]{natbib}
\usepackage{hyperref}
\usepackage{graphicx}
\usepackage{epstopdf}
\usepackage{bm}
\usepackage{epsfig}
\usepackage{graphics}
\usepackage{xspace}
\usepackage{amsfonts}
\usepackage{verbatim}
\usepackage{geometry}
\usepackage{color}
\usepackage[utf8]{inputenc}

\newcommand{\nn}{\nonumber}

\newcommand{\ep}{\epsilon}
\newcommand{\al}{\alpha}
\newcommand{\as}{\frac{\alpha_s C_F}{2\pi}}

\newcommand{\mL}{\mathcal{L}}

\newcommand{\bn}{\bar{n}}

\newcommand{\Tr}{\textrm{Tr}}
\newcommand{\half}{\frac{1}{2}}

\def\psl{p\!\!\!\slash}
\def\ksl{k\!\!\!\slash}
\def\nsl{n\!\!\!\slash}
\def\nbsl{\bar{n}\!\!\!\slash}
\def\Asl{A\!\!\!\slash}
\def\epsl{\ep^*\!\!\!\!\!\slash}

\def\epsm1{\frac{1}{\ep}}
\def\eps#1{\frac{1}{\ep^{#1}}}

\begin{document}

\begin{titlepage}

\vskip 25mm

\begin{center}
\Large\bf{Resummation of double logarithms  in loop-induced processes with effective field theory}
\end{center}

\vskip 8mm

\begin{center}

{\bf Jian Wang}\\
\vspace{10mm}
\textit{School of Physics, Shandong University, Jinan, Shandong 250100, China}\\
\vspace{5mm}

\end{center}

\vspace{10mm}


\begin{abstract}
The large  double logarithm in loop-induced processes is one kind of logarithm at subleading power,
which has a different origin from Sudakov double logarithms. 
We develop a method with soft-collinear effective theory to  resum these large double logarithms to all orders in 
the strong coupling constant.
\end{abstract}

\end{titlepage}


\section{Introduction}

To provide most precise theoretical predictions for observables at colliders,
it is helpful to resum various kinds of large logarithms to all orders in the coupling constants.
These large logarithms are usually induced by the soft and collinear radiations.
The resummation of these large logarithms has been achieved by making use of 
the factorization of the cross section to a set of functions at different energy scales
and the renormalization group evolutions controlled by the corresponding anomalous dimensions
\cite{Sterman:1986aj,Catani:1989ne,Korchemsky:1993uz,Contopanagos:1996nh,Forte:2002ni,Banfi:2004yd,Becher:2006nr,Luisoni:2015xha}.
So far, the formula and results at leading power have been extensively explored.
Beyond leading power, there are a number of  studies toward understanding the subleading
power threshold effects for colorless final states \cite{Bonocore:2014wua,Bonocore:2015esa,Bonocore:2016awd,DelDuca:2017twk,Bahjat-Abbas:2018hpv} or colored final states \cite{Bhattacharya:2018vph,vanBeekveld:2019prq,vanBeekveld:2019cks,Boughezal:2019ggi},
the subleading power corrections for $N$-jettiness   subtractions at next-to-next-to leading order in the strong coupling constant $\al_s$ \cite{Moult:2016fqy, Boughezal:2016zws,Moult:2017jsg,Boughezal:2018mvf,Ebert:2018lzn},  
the subleading power corrections  to the transverse momentum spectrum of a Higgs boson or a gauge boson
\cite{Balitsky:2017flc,Balitsky:2017gis,Ebert:2018gsn,Cieri:2019tfv},
the anomalous dimensions of subleading power operators \cite{Hill:2004if,Beneke:2005gs,Freedman:2014uta,Goerke:2017lei,Beneke:2017ztn,Beneke:2018rbh,Beneke:2019kgv},
and the resummation of double logarithms  at subleading power for the thrust observable  \cite{Moult:2018jjd,Moult:2019mog,Moult:2019uhz},
 the threshold cross section of Drell-Yan like processes  \cite{Beneke:2018gvs,Bahjat-Abbas:2019fqa,Beneke:2019mua,Beneke:2019oqx},
and the energy-energy correlator in the back-to-back limit \cite{Moult:2019vou}.

One kind of the large double logarithm at subleading power appears  in the loop-induced processes 
\cite{Jikia:1996bi,Fadin:1997sn,Kotsky:1997rq,Akhoury:2001mz,Melnikov:2016emg,Braaten:2017lxx,Braaten:2017ukc}, such
as the Higgs boson decay $H\to \gamma\gamma$ via a massive quark loop,
or similar processes with a massive quark propagator 
\cite{Penin:2014msa,Penin:2016wiw,Liu:2017vkm,Alte:2018nbn,Liu:2018czl,Alte:2019iug}.
If the mass of the quark in the loop, for example the bottom quark mass $m_b$, is much less than the Higgs boson mass $m_h$, 
the amplitude contains large double logarithms 
$\alpha_s^n\ln^{2n}( m_h^2/m_b^2 )$.
In contrast to the Sudakov double logarithms which are induced by soft and collinear gauge bosons,
these double logarithms are induced by a fermion.
For some specific processes, they have been obtained up to the two-loop level
\cite{Inoue:1994jq,Spira:1995rr,Fleischer:2004vb,Harlander:2005rq,Anastasiou:2006hc,Aglietti:2006tp,Spira:2016ztx},
and resummed to all orders 
by using the off-shell Sudakov form factor  \cite{Kotsky:1997rq,Akhoury:2001mz} 
or  by applying a sequence of identities graphically \cite{Liu:2017vkm,Liu:2018czl}.
In this work, we propose a method to resum the large double logarithms in loop induced processes
with an effective field  theory.
This study can help to understand the all order structure of subleading power logarithms,
and the method developed in this work may be also useful to resum general large logarithms at subleading power,
especially for the processes in which the subleading power corrections are numerically significant.
We show our resummation scheme through an example of the process $H\to \gamma\gamma$ via a bottom-quark loop,
leaving the generalization to non-Abelian cases to future work.
We notice that a different method has been developed in \cite{Liu:2019oav} to deal with the same problem.

\begin{figure}[h]
\centering
 \includegraphics[width=0.48\linewidth]{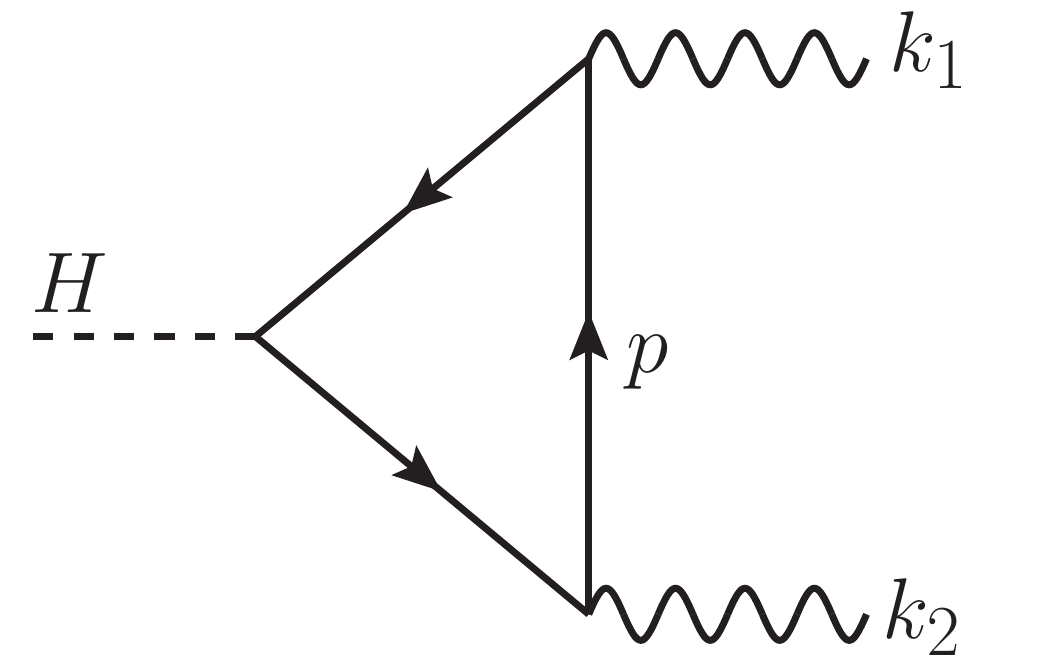}
  \caption{The leading order Feynman diagram of the decay $H\to \gamma\gamma$ via a bottom-quark loop.
  The other diagram with inverted fermion flow is not shown.} 
  \label{fig:sketch}
\end{figure}

\section{Factorization}

As shown in Fig.\ref{fig:sketch}, the leading order amplitude of $H \to \gamma(k_1)\gamma(k_2)$ via a bottom-quark loop  can be written as
\begin{align}
\mathcal{A} = &ie_q^2y_bN_c \int \frac{d^dp}{(2\pi)^d}
 \frac{\Tr[(\psl +m_b) \epsl~(k_2) (\psl+\ksl_2 +m_b) (\psl-\ksl_1+m_b)  \epsl~(k_1)]}{(p^2-m_b^2) [(p+k_2)^2-m_b^2]  [(p-k_1)^2-m_b^2]}
 \label{eqlo}
\end{align}
with $y_b$ the bottom quark's Yukawa coupling, its electric charge $e_q=\frac{-1}{3}e$ and $d=4-2\ep$.
It is convenient to choose two light-like directions $n$ and $\bar{n}$ such that
\begin{align}
k_1^{\mu} = \frac{m_h}{2}n^{\mu}, \qquad k_2^{\mu} = \frac{m_h}{2}\bar{n}^{\mu}.
\end{align}
Then, any momentum $q$ can be decomposed as
\begin{align}
q^{\mu} = q^+\frac{\bn^{\mu}}{2} + q^- \frac{n^{\mu}}{2} + q_{\perp}^{\mu}
\end{align}
with $q^+  \equiv q\cdot n,  q^-  \equiv q\cdot \bn.$
The $n$-collinear momentum scales as 
$(q^+,q^-,q_{\perp})\sim m_h (\lambda^2,1,\lambda)$,
while the soft momentum scales as
$(q^+,q^-,q_{\perp})\sim m_h(\lambda,\lambda,\lambda)$ with $\lambda=m_b/m_h$.
For later analyses, we also need hard-collinear momentum $m_h(\lambda, 1, \sqrt{\lambda})$
as well as quasi-hard-collinear momentum  $m_h(\lambda, 1, \lambda)$.
This quasi-hard-collinear momentum is present only for specific loop momentum.
It has the same offshellness as the hard-collinear momentum, but with smaller transverse momentum fluctuation.
The amplitude in Eq.(\ref{eqlo}) (or higher-order results) can be expanded in a series of $\lambda$
using the method of expansion by regions \cite{Smirnov:1997gx,Beneke:1997zp}.
However, we will  derive the leading contributions with soft-collinear effective theory (SCET)
\cite{Bauer:2000ew,Bauer:2000yr,Bauer:2001yt,Bauer:2002nz,Beneke:2002ph}.
Because the power counting is explicit in the effective Lagrangian and operators, 
we get a systematic control on the large logarithms  in power expansion and 
can resum them to all orders in $\alpha_s$.

First we present a factorization form of the amplitude to all orders in $\alpha_s$.
The QCD current that induces the process $H\to \gamma\gamma$ is given by
\begin{align}
O(x,y_1,y_2)= J(x) \cdot i \mL_{\gamma}(y_1) \cdot i\mL_{\gamma}(y_2)
\end{align}
where $J(x)\equiv -y_b \phi(x)\bar{\psi}(x) \psi (x)$ and $\mL_{\gamma}(y) \equiv e_q \bar{\psi}(y)\Asl(y) \psi(y) $
with $ \phi(x), \psi(x), A(x)  $ the Higgs, bottom quark, and photon field, respectively.
 The amplitude can be written as
\begin{align}
\mathcal{A}=&\langle k_1; k_2| \int d^d x \int d^d y_1 \int d^d y_2 \bold{T}\big[ O(x,y_1,y_2)\big]| p_h \rangle,
\label{eqamp}
\end{align}
where $k_1,k_2$ are the momenta of the photons,
and $p_h$ is the momentum of the Higgs boson.
We can eliminate the Higgs field and set $x=0$.
Now we match the QCD current to SCET,
\begin{align}
&O(y_1,y_2) \to  O_h(y_1,y_2) + 
\int ds dt [ \tilde{C}_n(s,t) O_n(y_1,y_2,s,t) \nn\\
&+\tilde{C}_{\bar{n}}(s,t) O_{\bar{n}}(y_1,y_2,s,t) 
+\tilde{C}_{s}(s,t) O_{s}(y_1,y_2,s,t) ].
\label{eqmatch}
\end{align}
Here $O_h(y_1,y_2) $ is obtained from $O(y_1,y_2)$ by neglecting the quark mass
and results in the amplitude $\mathcal{A}_H$.
The other terms are given by
\begin{subequations}
\begin{align} 
&\int d^d y_2 \bold{T}[O_n(y_1,y_2,s,t)] 
=\bold{T}[J^{B1}_n(s,t), i\mL_{m,n}^{(0)}(y_1) ] , \label{eq:coll}\\
&\int d^d y_1 \bold{T}[O_{\bar{n}}(y_1,y_2,s,t)] 
=\bold{T}[J^{B1}_{\bar n}(s,t), i \mL_{m,\bar{n}}^{(0)}(y_2) ] , \label{eq:anticoll}\\
&\bold{T}[O_s(y_1,y_2,s,t)]
=\bold{T}[J^{(A0,A0)}_{n\bar{n}}(s,t), i\mL_{s,n}^{(1/2)}(y_1),  i\mL_{s,\bar{n}}^{(1/2)}(y_2) ].\label{eq:soft}
\end{align}
\end{subequations}
The currents are defined following the convention in \cite{Beneke:2017ztn},
\begin{align}
J^{B1}_n(s,t)&=\frac{e_q y_b}{\bar{\mathcal{P}}_1}  \bar{\chi}_n(t\bar{n}) \epsl_{\perp}(k_2) \frac{\nbsl}{2}   \chi_n(s\bar{n}),
\label{eq:JB1}\\
J^{(A0,A0)}_{n\bar{n}}(s,t) &=-y_bJ^{A0}_{\bar{n}}(s) J^{A0}_{n}(t)=-y_b  \bar{\zeta}_{\bar{n}}(sn) \zeta_n(t\bar{n}),
\end{align}
where the collinear and hard-collinear quark field are \cite{Bauer:2002nz}
\begin{align}
\chi_n(x) \equiv W_c^{\dagger}\xi_c(x), ~~~~ \zeta_n(x) \equiv W_{hc}^{\dagger}\xi_{hc}(x),
\end{align}
and the operator $\bar{\mathcal{P}}_1$ picks out the $O(\lambda^0)$ momentum component of the  $\bar{\chi}_n $ field.
One of the inserted vertex can be found in \cite{Leibovich:2003jd}
\begin{align}
\mL_{m,n}^{(0)}(y) =\frac{e_q m_b m_h}{\bar{\mathcal{P}}_1\bar{\mathcal{P}}_2} \bar{\chi}_n(y)\Asl_{\perp}(y) \frac{\nbsl}{2}  \chi_n(y),
\end{align}
where we have omitted those terms that do not contribute to the amplitude.
The superscript of $\mL_{m,n}^{(0)}$ denotes that this is a leading power interaction.
The operator $\bar{\mathcal{P}}_{1,2}$ here act on the $\chi_n $ and $\bar{\chi}_n $ field, respectively.
The other inserted vertices are
\begin{align}
 \mL_{s,n}^{(1/2)}(y) =e_q \bar{\xi}_{\rm qhc}(y)\Asl_{\perp}(y)q_{s}(y) + {\rm h.c.},
 \label{eqLsn}
\end{align}
where we introduce the offshell field $ \bar{\xi}_{\rm qhc}(y)$ with the momentum $p_{\rm qhc}\sim (\lambda,1,\lambda)$
due to the momentum conservation.
Compared with the hard-collinear mode, it has smaller transverse momentum.
We  emphasise that this offshell field appears only  as an intermediated state.
The power counting of $ \mL_{s,n}^{(1/2)}$ is $O(\lambda^{1/2})$.
The appearance of the subleading current $J^{B1}$ in Eqs.(\ref{eq:coll},\ref{eq:anticoll}) 
and two inserted vertices $ \mL_{s,n}^{(1/2)}$   in Eq.(\ref{eq:soft})
indicates that the large double logarithms in this process are one kind of logarithms at subleading power.

The matching in Eq.(\ref{eqmatch}) can be obtained by integrating out hard momentum in the loop
or expanding the QCD amplitude with the method of expansion by regions,
and has been verified through reproducing the leading logarithms in fixed-order QCD results.

Then we define the hard function for $O_n$ as
\begin{align}
H_n(z)&=
\int ds dt e^{im_h z t} e^{im_h\bar{z}s} \tilde{C}_n(s,t) ,
\end{align}
where we have used $z$ to denote the momentum fraction of one jet  in all collinear final state
and $\bar{z}\equiv 1-z$.
The jet function for $O_n$ is defined as
\begin{align}
&\frac{-e_q m_b m_h N_c}{16\pi^2}\Tr[\epsl_{\perp}(k_1)\frac{\nsl}{2} \Gamma ] J_n(z) \equiv
\int d^d y
e^{ik_1\cdot y}
\langle k_1|  \bold{T}[ \bar{\chi}_{n,z}(0)\Gamma \chi_{n,\bar{z}}(0),  i \mL_{m,n}^{(0)}(y) ]| 0 \rangle  ,
 \label{eqJc}
\end{align}
where $\bar{\chi}_{n,z}(0)$ denotes the collinear jet field with a momentum fraction $z$ of the total collinear momentum, i.e., $p^-\equiv m_h z$, and $\Gamma$ represents any combination of Dirac matrices.
The amplitude induced by $O_n$ is given by
\begin{align}
\mathcal{A}_C=&\frac{y_b e_q^2N_c}{8\pi^2}\epsilon^*_{\perp}(k_1)\cdot \epsilon^*_{\perp}(k_2) m_b   \int_0^1 d z
\frac{1}{z}H_n (z)  J_n(z),
\label{eqAc}
\end{align}
where the factor $1/z$ comes from the denominator in Eq.(\ref{eq:JB1}).
Similarly we obtain the amplitude induced by $O_{\bar{n}}$, denoted as $\mathcal{A}_{\bar{C}}$.

The hard function for $O_s$ is given by
\begin{align}
H^s&=
\int ds dt e^{im_h s } e^{im_h t} \tilde{C}_s(s,t).
\end{align}
We define the jet function in the $n$-collinear direction
\begin{align}
    J_n^{s}(k_1,p)=\int d^d y e^{-ip\cdot y}\langle k_1|
    \bold{T}[\zeta(0),i\bar{\xi}_{\rm qhc}(y)\Asl_{\perp}(y)]|0\rangle
\end{align}
and in the $\bar{n}$-collinear direction
\begin{align}
    J_{\bar{n}}^{s}(k_2,\bar{p})=\int d^d y e^{i\bar{p}\cdot y}\langle k_2|
    \bold{T}[\bar{\zeta}(0),i\Asl_{\perp}(y)\xi_{\rm qhc}(y)]|0\rangle
    .
\end{align}
The soft function is then defined as
\begin{align}
 S^s(p,\bar{p})=&
\int d^{d}y_1 \int d^{d} y_2
e^{ip\cdot y_1 -i\bar{p}\cdot y_2 }
 \langle 0|  \bold{T}[ \hat{Y}^{\dagger}_{\bar{n}}(0,\bar{p})q_s(y_1)\bar{q}_s(y_2) \hat{Y}_{n}(0,p)]|0 \rangle ,
\label{eqSs}
\end{align}
where we have inserted 
the soft Wilson line along the hard-collinear particle,
\begin{align}
\hat{Y}_n(x,p)=\bold{\bar{P}} \exp\left( -i g_s \int^{\infty}_0 ds n\cdot A_s(x+s n) e^{-isn\cdot p}\right),
\end{align}
and the soft Wilson line along the hard-anti-collinear particle,
\begin{align}
\hat{Y}^{\dagger}_{\bar{n}}(x,\bar{p})=\bold{P} \exp\left( i g_s \int^{\infty}_0 ds \bar{n}\cdot A_s(x+s \bar{n}) e^{is\bar{n}\cdot \bar{p}}\right),
\end{align}
in order to decouple the soft interaction from the hard-(anti-)collinear particles.
These Wilson lines are obtained following the method in \cite{Chay:2004zn}.

To the leading logarithmic accuracy the jet function and soft function are given by
\begin{align}
     J_n^{s}(k_1,p)&=\epsl_{\perp}(k_1)\frac{\nsl}{2}\frac{1}{p^+}J_n^{s}(p^+),\\
    J_{\bar{n}}^{s}(k_2,\bar{p})&=\epsl_{\perp}(k_2)\frac{\nbsl}{2}\frac{1}{\bar{p}^-}J_{\bar{n}}^{s}(\bar{p}^-),\\
     S^s(p,\bar{p})&= \frac{im_b}{p^2-m_b^2}  (2\pi)^d \delta^{(d)}(p-\bar{p}) S^s(p),\label{eqSsll}
\end{align}
where the scalar functions $J_n^{s}(p^+),J_{\bar{n}}^{s}(\bar{p}^-),S^s(p)$ are just one at leading order in $\alpha_s$.
We keep the factor $m_b$ in Eq.(\ref{eqSsll}) because the helicity must be flipped on the soft quark propagator.
After contracting the Lorentz indices, 
we obtain the amplitude induced by $O_s$ 
\begin{align}
\mathcal{A}_S = & 2i y_b e_q^2 N_c \epsilon^*_{\perp}(k_1)\cdot \epsilon^*_{\perp}(k_2)  m_bH^s 
 \int \frac{d^{d}p}{(2\pi)^{d}}
 \frac{J_n^s(p^+)  S^s(p)  J_{\bar{n}}^s(\bar{p}^-)}{p^+ p^- (p^2 - m_b^2)}
 .
\label{eqAs1}
\end{align}

Summarising the above results,  we obtain the amplitude
\begin{align}
\label{eq0}
\mathcal{A}&
=(\mathcal{A}_H+\mathcal{A}_C+\mathcal{A}_{\bar{C}}- \mathcal{A}_S)[1+O(\lambda)].
\end{align}
The minus sign of $\mathcal{A}_S$ arises because 
the  zero-bin subtraction in the (anti-)collinear sectors has been performed 
\cite{Manohar:2006nz,Idilbi:2007ff,Idilbi:2007yi}.
The collinear, anti-collinear and soft sectors are separated by the rapidity of the momentum $p$. 
There exist rapidity divergences, as shown in Eqs.(\ref{eqAc},\ref{eqAs1}), 
which must be regularised in the intermediate steps but 
cancel in the end among the collinear, anti-collinear and soft sectors.

We choose the $\Delta$-regulator \cite{Chiu:2009yx} to regularise these rapidity divergences
\footnote{The analytic regulator \cite{Becher:2012xx} is often chosen to regularise the rapidity divergences.
However, it is not appropriate in this case of $H\to \gamma\gamma$
if we want to resum the large logarithms to all orders.
The reason is that 
the leading-order result contains poles like  $(p^-)^{-1-\alpha}$.
Higher-order loop corrections would generate  structures like $(p^-)^{-\ep}/\ep$.
As a result, the amplitude at higher orders contains $(p^-)^{-1-\alpha-\ep}/\ep$.
One needs to expand the result first  in $\alpha$ and then in $\ep$ to get the correct result.
This means that one can not perform renormalization simply before integrating over $p$.
But after integrating over $p$, it is not easy to distinguish different origins of the large logarithms,
i.e., the factorization structure becomes unclear.
}. 
Accordingly, we take the replacement of the denominators,
\begin{align}
\frac{1}{ (p-k_1)^2-m_b^2 }& \to \frac{1}{ (p-k_1)^2-m_b^2 +\Delta_1}, \\ 
\frac{1}{ (p+k_2)^2-m_b^2 }& \to \frac{1}{ (p+k_2)^2-m_b^2 +\Delta_2}. 
\end{align}
These regulators $\Delta_{1,2}$ have mass dimension two,
and are assumed to be much less than $m_b^2$ but can not be dropped in the denominator
even after power expansion. 
Therefore we rewrite Eqs.(\ref{eqAc},\ref{eqAs1}) as
\begin{align}
\mathcal{A}_C=&\frac{y_b e_q^2N_c}{8\pi^2} \epsilon^*_{\perp}(k_1)\cdot \epsilon^*_{\perp}(k_2)  m_b   
\int_0^1 d z
\frac{1}{z+\Delta_2/m_h^2}H_n (z)  J_n(z).
\label{eqAc2}
\end{align}
and
\begin{align}
\mathcal{A}_S = & 2i y_b e_q^2 N_c  \epsilon^*_{\perp}(k_1)\cdot \epsilon^*_{\perp}(k_2) m_bH^s\nn\\ &
 \int \frac{d^{d}p}{(2\pi)^{d}}
 \frac{J_n^s(p^+)  S^s(p)  J_{\bar{n}}^s(\bar{p}^-)}{(p^+ - \Delta_1/m_h) (p^-+\Delta_2/m_h) (p^2 - m_b^2)} . \label{eqAs}
\end{align}
Notice that the $\Delta$-regulator is only applied for the integration of the outmost quark loop momentum. 
It is not implemented for those higher-order loop integration induced by gluons. 
Therefore, with this regulator, the leading order rapidity divergences exist in the form of $\ln^n \Delta_{1,2}/m_h^2$,
while higher-order divergences are still in the form of $1/\epsilon^n$. 
The large logarithms associating with these higher-order divergences can be separated from the leading order ones without ambiguity
since they are in different form now. 

Besides the rapidity divergence, there are usual infrared and ultraviolet  divergences
in the hard, collinear and soft sectors, respectively. 
They can be tamed with dimensional regularisation.

The loop-induced processes are different from those having tree-level contributions, since
the leading order contributions in the effective theory, i.e., the hard, collinear and soft sectors, already contain divergences.
These leading order divergences are not renormalized as usual in the multiplicative renormalization scheme.
Instead, they cancel each other among the hard, collinear and  soft sectors.

To see the structure of divergences more clearly, 
we rewrite Eq.(\ref{eq0}) as  (dropping the $O(\lambda)$ corrections)
\begin{align}
\mathcal{A}=&\mathcal{A}_H\left(\ep\right)
+\mathcal{A}_{C}\left(\ep,\Delta_2\right)
+\mathcal{A}_{\bar{C}}\left(\ep,\Delta_1\right)
-\mathcal{A}_S\left(\ep,\Delta_1,\Delta_2\right) .
\end{align}
The $\ep$-poles in $\mathcal{A}_H$ are infrared divergences since the propagators are all massless.
The $\ep$-poles in $\mathcal{A}_{C},\mathcal{A}_{\bar{C}}$ and $\mathcal{A}_S$
are ultraviolet divergences, generated when the transverse momentum $p_T$ is integrated up to infinity.
Then we can rearrange the above equation as
\begin{align}
\mathcal{A}&=\mathcal{A}_H\left(\ep\right)
+\left[\mathcal{A}_{C}\left(\ep,\Delta_2\right)\right]_{p_T > m_h}
+\left[\mathcal{A}_{\bar{C}}\left(\ep,\Delta_1\right)\right]_{p_T > m_h}
-\left[\mathcal{A}_S\left(\ep,\Delta_1,\Delta_2\right)\right]_{p_T > m_h}\nn\\
&+\left[\mathcal{A}_{C}\left(\Delta_2\right)\right]_{p_T \le m_h}
+\left[\mathcal{A}_{\bar{C}}\left(\Delta_1\right)\right]_{p_T \le m_h} 
-\left[\mathcal{A}_S\left(\Delta_1,\Delta_2\right)\right]_{p_T \le m_h}.
\label{eq:splitpt}
\end{align}
Notice again that we divide  only the transverse momentum integration for the outmost quark loop.
The pieces in the first line contain $1/\ep^n$ poles, which cancel each other, while
the pieces in the second line contain no such poles and thus finite.
The cancellation of $1/\ep^n$ poles is guaranteed to all orders in $\alpha_s$ because  there is no
such divergences on the left-hand side of this equation 
and the right-hand side is a complete leading power expansion. 
In fact, one can consider the division of the integration range of $p_T$ as a way of renormalization
and the cutoff $m_h$ is the renormalization scale. 
It is possible to choose another renormalization scale without changing the final result.
Since the intrinsic scale of $\mathcal{A}_H$ is $m_h$,   
setting $m_h$ as the cutoff scale could make the sum of the first line not contain any logarithms. 
The cancellation of the $\Delta$-regulators in the last line holds to all orders in $\alpha_s$.
This is because the left-hand side does not depend on these regulators,
and the regulators exist only in the $p^\pm$ integration, rather than the $p_T$ integration,
similarly to the situation in transverse momentum resummation.
Therefore, in each fixed value of $p_T$, the $\Delta$-dependences  cancel out.
Moreover, the pieces in the first line have a single intrinsic scale $m_h$ and therefore 
generate no large logarithms. 
So  we can neglect these contributions if we want to study only large logarithms
\footnote{We have checked that the poles in this part cancel out up to two-loop level. See the appendix.}.
In the following we use a subscript $R$ to denote the quantities with the constraint $p_T\le m_h$.

Each piece in the second  line receives QCD corrections at higher orders,
but can be renormalized as usual, since the leading order is finite now.
We will show the next-to-leading (NLO) QCD corrections in the following section.

\section{NLO corrections}
\label{sec:nlo}

We first consider the contribution from the $O_n$ current.
 The NLO hard function $H_n$ can be obtained by calculating the 
one-loop corrections,
\begin{align}
&H_n(z) =1 - \frac{\alpha_s C_F }{2\pi} 
 \ln z\left( \frac{1}{\ep}- \ln \frac{-m_h^2}{\mu^2}
-\half \ln z  \right) .
\end{align}
We show only the double logarithms in the result.
The above result arises from the expansion of 
$$\frac{1}{\ep^2}\bigg[ \bigg(\frac{-m_h^2}{\mu^2} \bigg)^{-\ep} - \bigg(\frac{-m_h^2 z}{\mu^2} \bigg)^{-\ep} \bigg] .$$
The two scales reflect the fact that there are two collinear jet fields in the collinear direction, 
which is a feature of subleading power operators.
By power counting, $m_h^2$ and $m_h^2 z$ are of the same scale, i.e. $O(1)$.

From the definition  in  Eq.(\ref{eqJc}),
we obtain the NLO jet function,
\begin{align}
&J_{n,R} (z) =\ln \frac{m_h^2}{m_b^2} 
\bigg\{ 1+
\frac{\alpha_s C_F}{2\pi} \ln z 
 \bigg[ \frac{ 1}{ \ep} 
-  \ln \frac{m_h^2}{\mu^2} 
+ \half \ln z +\half \ln  \frac{m_h^2}{m_b^2} \bigg]\bigg\}.
\end{align}
As shown in Eq.(\ref{eqJc}), the jet function is a function of $z$ and the bottom quark's mass $m_b$.
The scale $m_h$ appears because we use the cutoff renormalization scheme.
Before the integration over  $p_T$, we can see that the intrinsic jet scale is of order $\sqrt{p_T^2+m_b^2}$.
Since the corresponding hard function is insensitive to the transverse momentum of the external particles
of the operators, we need to integrate over $p_T$.
As a result, the resulting scale dependence is in such a complicated form. 
After performing the convolution between the hard and jet functions in Eq.(\ref{eqAc2}), 
we obtain the amplitude induced by $O_n$,
\begin{align}
\mathcal{A}_{C,R}
&=\frac{y_b  e_q^2}{8\pi^{2}} N_c m_b \epsilon^*_{\perp}(k_1)\cdot \epsilon^*_{\perp}(k_2) 
\bigg\{
\ln \frac{m_h^2}{\Delta_2}\ln \frac{m_h^2}{m_b^2}\\&
-\frac{\alpha_s C_F }{2\pi}
\left[ \frac{1}{3}\ln^3 \frac{\Delta_2}{m_h^2} \ln \frac{m_h^2}{m_b^2}
+ \frac{1}{4} \ln^2\frac{\Delta_2}{m_h^2}\ln ^2 \frac{-m_h^2}{m_b^2} \right] \bigg\},\nn
\end{align}
where we have kept only the leading logarithms.
We  see that the $1/\ep$-poles and $\mu$ scale dependent terms, which are induced by higher-order gluon loops, cancel
between the hard and jet function in this sector.
Similarly, we get $\mathcal{A}_{\bar{C},R}$ by replacing $\Delta_2 \to \Delta_1$.

Then we  consider the contribution from the $O_s$ current.
The hard function in this sector is straightforward to calculate,
\begin{align}
H^s=
1-\frac{\alpha_s C_F }{2\pi} 
\left[ \frac{ 1}{\ep^2}
-\frac{\ln (-m_h^2/\mu^2)}{\ep}+\half \ln^2 \frac{-m_h^2}{\mu^2} \right] .
\label{eq:Hs}
\end{align}
From the definitions, we can also calculate the NLO jet functions
\begin{align}
J^s_n(p^+)=&1+
\frac{\alpha_s C_F }{2\pi} 
 \left[ \frac{ 1}{\ep^2}-\frac{\ln (m_hp^+/\mu^2)}{\ep}+\half \ln^2 \frac{m_hp^+}{\mu^2}   \right]
, \\
J^s_{\bar{n}}(\bar{p}^-)=&  1
+\frac{\alpha_s C_F }{2\pi}
 \left[ \frac{ 1}{\ep^2}-\frac{\ln (-m_h\bar{p}^-/\mu^2)}{\ep}+\half \ln^2 \frac{-m_h\bar{p}^-}{\mu^2}   \right]
,
\end{align}
and the NLO soft function
\begin{align}
S^s(p)=&
1-\frac{\alpha_s C_F }{2\pi}
 \left[ \frac{ 1}{\ep^2}-\frac{\ln (p^+p^-/\mu^2)}{\ep}+\half \ln^2 \frac{p^+p^-}{\mu^2}  \right] .
\label{eq:Ss}
\end{align}
Notice that these jet and soft functions do not contain any plus distributions.
This is due to our choice of regulators (see Eq.(\ref{eqAs})),
which makes the integrations over $p^+$ and $p^-$ well-defined even if $p^+\to 0$ or $p^-\to 0$.

Inserting the hard, jet and soft functions to Eq.(\ref{eqAs}), 
we obtain the amplitude induced by $O_s$,
\begin{align}
&\mathcal{A}_{S,R}  = \frac{y_b e_q^2}{8\pi^2}N_c m_b \epsilon^*_{\perp}(k_1)\cdot \epsilon^*_{\perp}(k_2) \nn\\&
\times \bigg\{
\ln\frac{m_h^2}{\Delta_1}\ln\frac{m_h^2}{m_b^2}
+\ln\frac{m_h^2}{\Delta_2}\ln\frac{m_h^2}{m_b^2}-\half L^2 \nn\\&
-\frac{\alpha_s C_F}{2\pi}
\bigg[ \bigg( \frac{1}{3}\ln^3 \frac{\Delta_2}{m_h^2} \ln \frac{m_h^2}{m_b^2}
+ \frac{1}{4} \ln^2\frac{\Delta_2}{m_h^2}\ln ^2 \frac{-m_h^2}{m_b^2} \bigg) \nn \\
&+\bigg(  \frac{1}{3}\ln^3 \frac{\Delta_1}{m_h^2} \ln \frac{m_h^2}{m_b^2}
+ \frac{1}{4} \ln^2\frac{\Delta_1}{m_h^2}\ln ^2 \frac{-m_h^2}{m_b^2}
-\frac{1}{24}L^4 \bigg)\bigg]\bigg\}
\label{eq:AsR}
\end{align}
with $L\equiv \ln( -m_h^2/m_b^2-i0)$.
Once again, we find that the poles and scale dependent terms from the gluon loops  cancel.

Combining the above contributions from the (anti-)collinear and soft sectors, we obtain
\begin{align}
&\mathcal{A}_{C,R} +\mathcal{A}_{\bar{C},R} - \mathcal{A}_{S,R}
=\frac{ y_b e_q^2   }{8\pi^2  } N_c  m_b \epsilon^*_{\perp}(k_1)\cdot \epsilon^*_{\perp}(k_2) 
\bigg( \half L^2-
\frac{\alpha_s C_F}{2\pi}\frac{1}{24}
L^4
+O(\alpha_s^2)
\bigg).
\end{align}
All the dependence on the regulators $\Delta_1$ and $\Delta_2$ cancels,
and we reproduce the leading large logarithms of the QCD result up to the second order in $\al_s$
\cite{Fleischer:2004vb,Harlander:2005rq,Aglietti:2006tp}.

Another important observation is that  $\mathcal{A}_{C,R} \propto \ln\frac{m_h^2}{\Delta_2} $
because of the $\Delta$-regulator in Eq.(\ref{eqAc2}).
Similar argument shows that $\mathcal{A}_{\bar{C},R} \propto \ln\frac{m_h^2}{\Delta_1}$.
Therefore, after evaluating the integrations, 
we can set $\Delta_{1,2}=m_h^2$ so that $\mathcal{A}_{C,R} =\mathcal{A}_{\bar{C},R}=0$ and 
that the final result gets contribution just from $\mathcal{A}_{S,R} $.
This feature indicates that we need to analyze only the soft sector to resum the large double logarithms.

\section{Resummation}

From the above analysis, we have found that the main task is to calculate the result in the soft sector $\mathcal{A}_{S,R}\left(\Delta_1,\Delta_2\right)$.
All the logarithms in the hard, jet and soft functions of this sector
are scale dependent, and therefore can be resummed by using the corresponding renormalization
group evolution equations.
From the NLO results for the bare functions given in Eqs.(\ref{eq:Hs}-\ref{eq:Ss}), 
we derive the renormalization group evolution equations of the renormalized functions in the $\overline{\rm MS}$ scheme,
\begin{align}
\frac{dH^s(\mu)}{d\ln \mu} & = \bigg[ C_F \gamma_{\rm cusp} \ln \frac{-m_h^2}{\mu^2} \bigg]H^s(\mu),
\label{eq:h}\\
\frac{dJ_n^s(\mu)}{d\ln \mu} & = \bigg[ -C_F \gamma_{\rm cusp} \ln \frac{m_hp^+}{\mu^2} \bigg]J_n^s(\mu),\\
\frac{dJ_{\bar{n}}^s(\mu)}{d\ln \mu} & = \bigg[- C_F \gamma_{\rm cusp} \ln \frac{-m_h p^-}{\mu^2} \bigg]J^s_{\bar{n}}(\mu),\\
\frac{dS^s(\mu)}{d\ln \mu} & = \bigg[ C_F \gamma_{\rm cusp} \ln \frac{p^+ p^-}{\mu^2} \bigg]S^s(\mu),
\label{eq:s}
\end{align} 
where $\gamma_{\rm cusp}$ is the cusp anomalous dimension \cite{Becher:2009qa}.
It is evident that 
\begin{align}
\frac{d\ln [H^s(\mu)J_n^s(\mu)S^s(\mu)J^s_{\bar{n}}(\mu)] }{d\ln \mu}=0.
\end{align}
Solving the evolution equations in Eqs.(\ref{eq:h}-\ref{eq:s}), we get
\begin{align}
\label{eq15}
H^s(\mu)J_n^s(\mu)S^s(\mu)J^s_{\bar{n}}(\mu)&=
\exp\big[2C_F\big(S(\mu_h,\mu)-S(\mu_c,\mu)-S(\mu_{\bar{c}},\mu)+S(\mu_s,\mu) \big)\big] \nn \\
&=
 \exp \bigg[ -\frac{\alpha_s C_F}{2\pi} \ln \frac{-p^+}{m_h} \ln \frac{p^-}{m_h} +O(\alpha_s^2) \bigg],
\end{align}
where the  function $S(\mu_i,\mu_f)$ is defined by \cite{Neubert:2004dd}
\begin{align}
S(\nu,\mu) =-\int_{\alpha_s(\nu)}^{\alpha_s(\mu)}d\alpha \frac{\gamma_{\rm cusp}}{\beta(\alpha)} 
\int_{\alpha_s(\nu)}^{\alpha}\frac{d\alpha'}{\beta(\alpha')}.
\end{align}
In the last line, we have neglected those terms at $O(\alpha_s^2)$, which contribute at most $\alpha_s^2 L^3$.
We have chosen the typical scales to be $\mu_h^2=-m_h^2, \mu_{c}^2=m_hp^+,\mu_{\bar{c}}^2=-m_hp^-,\mu_s^2=p^+ p^-$ as indicated in the fixed order calculation,
though their explicit values do not affect the final result.

\begin{figure}[htbp]
\begin{center}
 \includegraphics[width=0.48\linewidth]{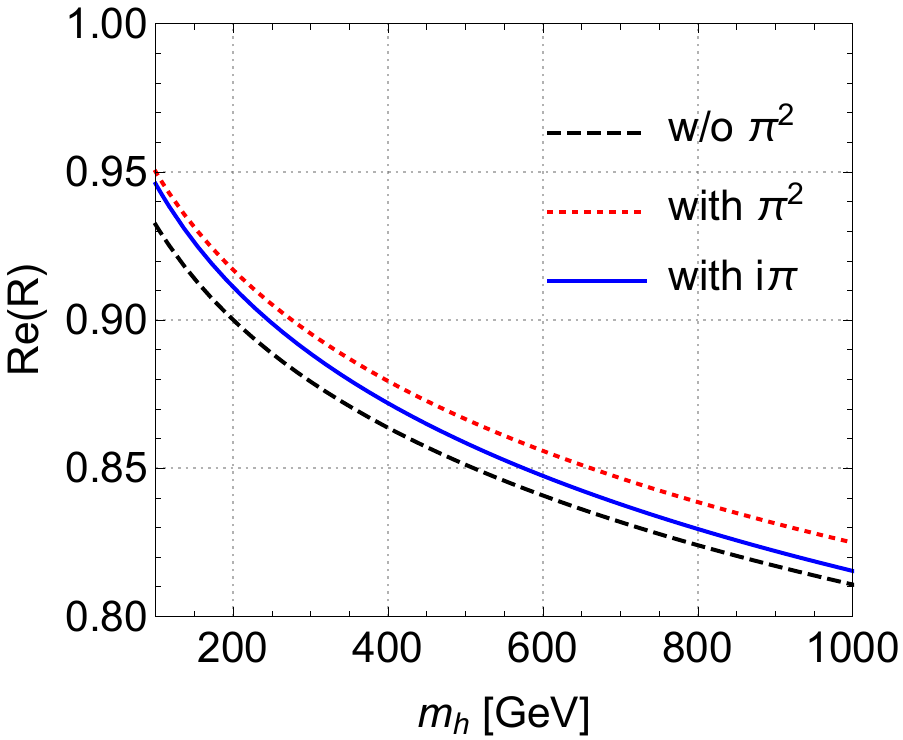}
  \includegraphics[width=0.48\linewidth]{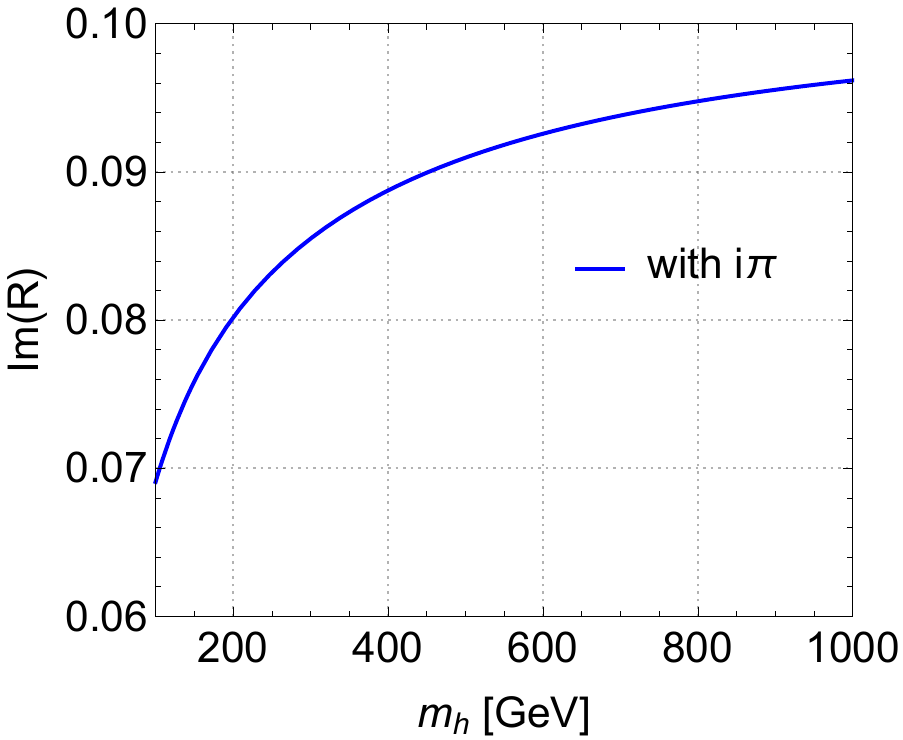}
 \\
 \caption{The ratio $R$ of the resummed result over the leading order result.
 Its real and imaginary part are shown on the left and right plot, respectively.
The dashed black and dotted red line (on the left plot) denotes the result of Eq.(\ref{eq16}) 
with $L^2$ replaced by $\ln^2(m_h^2/m_b^2)$ and $\ln^2(m_h^2/m_b^2)-\pi^2$, respectively. 
The blue  line represents the 
result with $L^2=\ln^2(m_h^2/m_b^2)-\pi^2-2i\pi \ln (m_h^2/m_b^2)$.
We have used $\alpha_s(m_h)=0.113, m_b=5 $ GeV.
}
\label{fig2}
\end{center}
\end{figure}

\begin{figure}[htbp]
\begin{center}
 \includegraphics[width=0.45\linewidth]{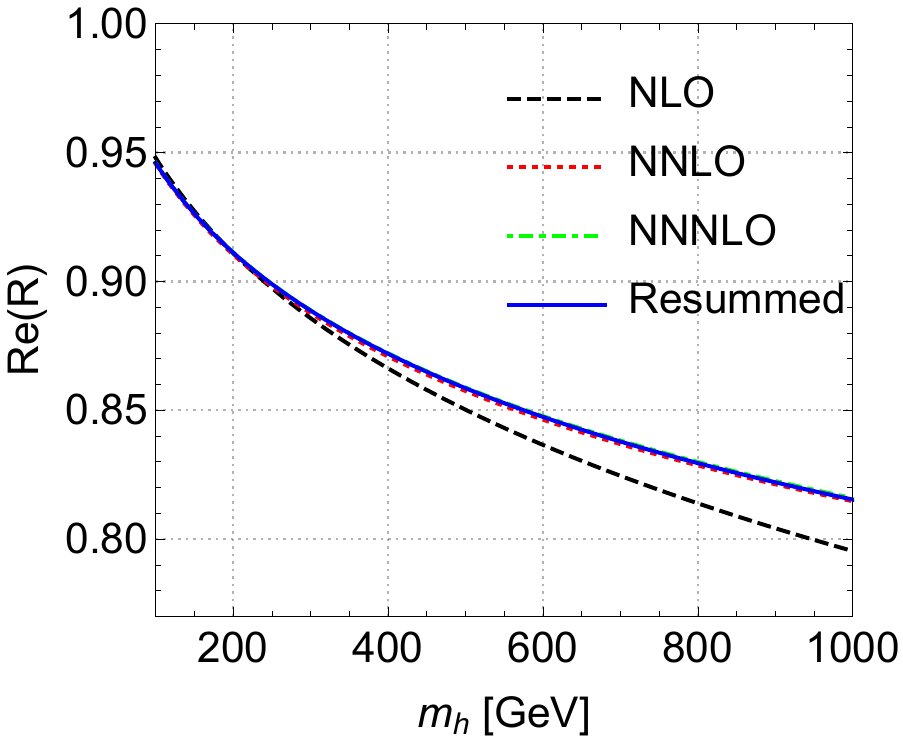}
  \includegraphics[width=0.45\linewidth]{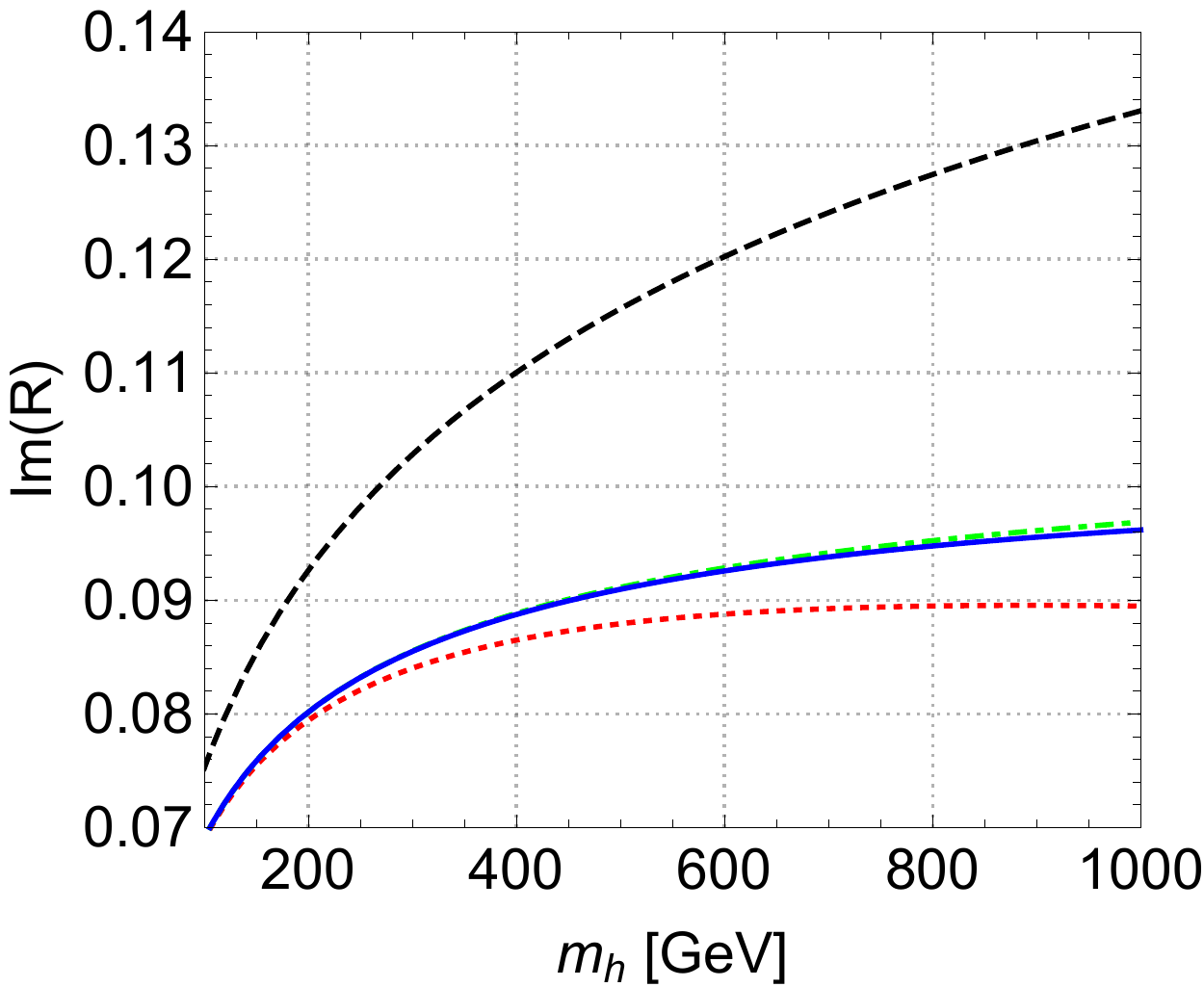}
 \\
 \caption{Same as Fig. \ref{fig2} with $L^2=\ln^2(m_h^2/m_b^2)-\pi^2-2i\pi \ln (m_h^2/m_b^2)$ but with more lines to show the expanded results.
 Notice that the legends NLO, NNLO, NNNLO do not denote the full fixed-order results, but only the leading logarithms. 
}
\label{fig3}
\end{center}
\end{figure}

Inserting Eq.(\ref{eq15}) in Eq.(\ref{eqAs}), we can perform the integrations over $p$,
while keeping $\Delta_{1,2}$ as small regulators,  to obtain the result of $\mathcal{A}_{S,R}(\Delta_1,\Delta_2)$ to all orders in $\al_s$;
the first two orders have been given in Eq.(\ref{eq:AsR}).
Then we set $\Delta_{1,2}=m_h^2$ so that $\mathcal{A}_{C,R} $ and $\mathcal{A}_{\bar{C},R} $ are vanishing.
As a consequence, we  obtain
\begin{align}
& \mathcal{A}_{C,R}  + \mathcal{A}_{\bar{C},R} -\mathcal{A}_{S,R}
 =  \frac{ y_b e_q^2  }{8 \pi^2 } N_c m_b \epsilon^*_{\perp}(k_1)\cdot \epsilon^*_{\perp}(k_2)  \half L^2
\times {}_{2}F_{2}(1,1;\frac{3}{2},2; -\frac{\alpha_s C_F}{8\pi}L^2 ),
 \label{eq16}
\end{align}
which agrees with 
the result in Ref.\cite{Akhoury:2001mz}, except that we have reproduced the double logarithms 
in the form $ \ln^2 (-m_h^2/m_b^2-i0) $, instead of $ \ln^2(m_h^2/m_b^2) $.
This is because we have considered the hard and collinear sectors besides the soft sector.
As a consequence,  it allows us to resum the large $\pi^2$ terms in the perturbative calculations too.
The generalized hypergeometric function ${}_{2}F_{2}(1,1;\frac{3}{2},2; -x)$ looks strange but actually has an exponential structure
$\sqrt{\pi}x^{3/2}e^{-x}/2$
at $x\to \infty$.
The impact of the resummed result in Eq.(\ref{eq16}) are shown in Fig.\ref{fig2} and Fig.\ref{fig3}.
For the standard model value $m_h=125$ GeV, the ratio of the resummed result over the leading order result is 
$0.935+0.073i$.
The impact becomes more significant if the Higgs boson mass $m_h$ increases.
For comparison, the result of Ref.\cite{Akhoury:2001mz} is shown in the black dashed line in Fig.\ref{fig2}.
It differs from the real part of our result by less than $2\%$, 
but it contains no imaginary part.
We also show the comparison with the result in which the $\pi^2$ terms are kept, i.e.,
replacing $L^2$ by $\ln^2(m_h^2/m_b^2)-\pi^2$.
We see that the difference is tiny around $m_h=125$ GeV, growing to about $2\%$ around $m_h=1000$ GeV.
In Fig.\ref{fig3}, we expand the resummed result to the first few orders.
The real part converges quickly since the contribution from the first three orders (NNLO) already overlaps
with the resummed result over a large range of $m_h$.
The imaginary part converges slower.
But the sum of the first four orders is already a good approximation of the resummed result.

\section{Conclusions}

We have provided a method to resum the large logarithms in loop-induced processes 
with soft-collinear effective theory. 
This method is different from the conventional threshold or transverse momentum resummation, 
because the leading-order contributions have already
divergences in the soft and collinear limits.
By adopting a $\Delta$-regulator and a cut-off renormalization for the quark loop transverse momentum,
we can make the leading-order result finite.
Further, there are several sectors in the effective field theory contributing to the 
leading power expansion of the QCD amplitude. 
Each of them  depends on the rapidity regulators. 
In particular, the contributions from (anti-)collinear sectors are proportional to $\ln (\Delta_{1,2}/m_h^2)$ 
so that we can choose a special value of the regulator to make them vanishing.
Then one only needs to consider the contribution from the soft sector,
which has a structure ready to be resummed.
Expanding the resummed result to the first two orders, we find agreement with  previous QCD calculations.
Compared with the resummation method using off-shell Sudakov form factor,
we reproduce the full double logarithms $\ln ^2 (-m_h^2/m_b^2-i0)$.
As a consequence, the large $\pi^2$ terms and the leading contribution of the imaginary part of the amplitude 
can also be resummed. 
In future, it would be interesting to explore the resummation beyond  leading logarithmic accuracy.
It is also promising to extend our scheme to more general cases, 
such as processes with non-Abelian gauge bosons  or 
processes with more external particles, 
which are more important for collider phenomenology. 


\section*{Acknowledgments}
We appreciate Matthias Neubert's suggestion of this project and many inspiring comments very much.
We would like to thank Ze Long Liu and Ben D. Pecjak  for useful discussions.
The work of JW has  been supported  by the BMBF project No. 05H15WOCAA and 05H18WOCA1  when he was in Technische Universit\"at M\"unchen and by the program for Taishan scholars.

\appendix
\section{Cancellation of the divergences at the two-loop level}
\label{app:twoloop}
We have claimed in the main text that the divergences in the first line of Eq.(\ref{eq:splitpt}) cancel.
Here we present the explicit results at two loop.
Since the divergences will appear after integrating $p_T>m_h$, we must keep the $(p_T^2+m_b^2)^{-\ep}$ at higher orders.
Firstly, in the collinear sector, the higher-order hard and collider corrections are given by 
\begin{align}
B_H & =-\as \eps{2} \bigg[ \bigg(\frac{-m_h^2}{\mu^2} \bigg)^{-\ep} - \bigg(\frac{-m_h^2 z}{\mu^2} \bigg)^{-\ep} \bigg] ,\\
B_C  & =-\as \eps{2} \bigg[ \bigg(\frac{m_T^2}{\mu^2} \bigg)^{-\ep} - \bigg(\frac{m_T^2 }{z\mu^2} \bigg)^{-\ep} \bigg]
\end{align}
with $m_T^2\equiv p_T^2+m_b^2$. If we expand the above result in $\ep$, we obtain the $O(\alpha_s)$ part in Sec.\ref{sec:nlo}.
Here we prefer to keep the full $\ep$ dependence because all the power series of $\ep$ will contribute to the divergence after integration.
We have kept the $p_T$ dependence in $B_C$ in order to see that the cancellation of $\Delta$ regulators and divergences 
among the (anti-)collinear and soft sectors more clearly.
It is ready to perform the  integration,
\begin{align}
&\mu^{2\ep}\int_{m_h^2}^{\infty}dm_T^2 (m_T^2)^{-1-\ep} \int_0^1 dz\frac{1}{z+\delta_2}(B_H+B_C) 
= \mu^{2\ep}\int_{m_h^2}^{\infty}dm_T^2 (m_T^2)^{-1-\ep}\nn\\
& \as \eps{2}
\left[ 
\left( \frac{-m_h^2}{\mu^2} \right)^{-\ep} \left( -\epsm1 (1-\delta_2^{-\ep})+\ln \delta_2 \right) 
+\left( \frac{m_T^2}{\mu^2} \right)^{-\ep} \left( \epsm1 (1-\delta_2^{\ep})+\ln \delta_2 \right)
\right]\nn\\
=& \as \left( \frac{1}{4\ep^2}\ln^2\delta_2 -  \frac{1}{4\ep}\ln^3\delta_2 \right)+O(\ep^0)
\label{eq:bhbc}
\end{align}
with $\delta_2 \equiv \Delta_2/m_h^2$. As before, we only keep the leading logarithms in the calculation.
The anti-collinear result is obtained by changing $\delta_2\to \delta_1\equiv \Delta_1/m_h^2$. 

Then we turn to the soft sector. The higher-order hard, (anti-)collinear and soft corrections are given by 
\begin{align}
D_H & =-\as \eps{2} \bigg( \frac{-m_h^2}{\mu^2} \bigg)^{-\ep}  , \\
D_C  & =\as \eps{2}  \bigg(\frac{m_T^2 }{z\mu^2} \bigg)^{-\ep} , \\
D_{\bar{C}} & =\as \eps{2}   \bigg(\frac{-m_h^2 z}{\mu^2} \bigg)^{-\ep},\\
D_S  & =-\as \eps{2}  \bigg(\frac{m_T^2}{\mu^2} \bigg)^{-\ep} .
\end{align}
We see that their sum $D_t\equiv D_H+D_C  + D_{\bar{C}} +D_S$ is the same as the collinear sector $B_H+B_C$.
We divide the integration of $z$ into two parts,
\begin{align}
\mu^{2\ep}\int_{m_h^2}^{\infty}dm_T^2 (m_T^2)^{-\ep} \bigg( \int_0^1 dz +\int_1^{\infty} dz \bigg)  \frac{D_t}{z+\delta_2} \frac{1}{m_T^2-z\Delta_1}.
\end{align}
In the first part, we can drop the $\Delta_1$ regulator since $z$ is finite in this case.
In the second part, we can drop the $\delta_2$ regulator since $z$ is larger than 1.
As a consequence, the first part after integration will give the same result as Eq.(\ref{eq:bhbc}).
The second part is given by
\begin{align}
& \mu^{2\ep}\int_{m_h^2}^{\infty}dm_T^2 (m_T^2)^{-1-\ep}\as \eps{2} 
\Bigg\{ 
\left( \frac{-m_h^2}{\mu^2} \right)^{-\ep} \left[ \epsm1 \left(1-\left(\frac{-\Delta_1}{m_T^2}\right)^{\ep}\right)
+\ln \frac{-\Delta_1}{m_T^2} \right] \nn\\
&\hspace{4.5cm}+\left( \frac{m_T^2}{\mu^2} \right)^{-\ep} \left[ -\epsm1 \left(1-\left(\frac{-\Delta_1}{m_T^2}\right)^{-\ep}\right)
+\ln \frac{-\Delta_1}{m_T^2} \right]
\Bigg\} \nn \\
=&\as \left( \frac{-m_h^2}{\mu^2} \right)^{-2\ep}
\bigg(  \frac{-1}{4\ep^4}  + \frac{1}{4\ep^2}\ln^2\delta_1 -  \frac{1}{4\ep}\ln^3\delta_1\bigg)+O(\ep^0).
\end{align}
The $\delta_1$ dependent part has the same form as  Eq.(\ref{eq:bhbc}).

Lastly, the hard sector is given by
\begin{align}
\as \left( \frac{-m_h^2}{\mu^2} \right)^{-2\ep}  \frac{-1}{4\ep^4} .
\end{align}
And we see immediately that the divergences in 
\begin{align}
\mathcal{A}_H\left(\ep\right)
+\left[\mathcal{A}_{C}\left(\ep,\Delta_2\right)\right]_{p_T > m_h}
+\left[\mathcal{A}_{\bar{C}}\left(\ep,\Delta_1\right)\right]_{p_T > m_h}
-\left[\mathcal{A}_S\left(\ep,\Delta_1,\Delta_2\right)\right]_{p_T > m_h}
\end{align}
cancel each other.
Moreover, based on the above results, one can also find that
the $\Delta$ dependences in $\mathcal{A}_{C}+\mathcal{A}_{\bar{C}}-\mathcal{A}_S$ cancel for each fixed $p_T$.


\bibliography{Res}

\providecommand{\href}[2]{#2}\begingroup\raggedright\begin{thebibliography}{10}

\bibitem{Sterman:1986aj}
G.~F. Sterman, \emph{{Summation of Large Corrections to Short Distance Hadronic
  Cross-Sections}},
  \href{https://doi.org/10.1016/0550-3213(87)90258-6}{\emph{Nucl. Phys.}
  {\bfseries B281} (1987) 310}.

\bibitem{Catani:1989ne}
S.~Catani and L.~Trentadue, \emph{{Resummation of the QCD Perturbative Series
  for Hard Processes}},
  \href{https://doi.org/10.1016/0550-3213(89)90273-3}{\emph{Nucl. Phys.}
  {\bfseries B327} (1989) 323}.

\bibitem{Korchemsky:1993uz}
G.~P. Korchemsky and G.~Marchesini, \emph{{Resummation of large infrared
  corrections using Wilson loops}},
  \href{https://doi.org/10.1016/0370-2693(93)90015-A}{\emph{Phys. Lett.}
  {\bfseries B313} (1993) 433}.

\bibitem{Contopanagos:1996nh}
H.~Contopanagos, E.~Laenen and G.~F. Sterman, \emph{{Sudakov factorization and
  resummation}},
  \href{https://doi.org/10.1016/S0550-3213(96)00567-6}{\emph{Nucl. Phys.}
  {\bfseries B484} (1997) 303}
  [\href{https://arxiv.org/abs/hep-ph/9604313}{{\ttfamily hep-ph/9604313}}].

\bibitem{Forte:2002ni}
S.~Forte and G.~Ridolfi, \emph{{Renormalization group approach to soft gluon
  resummation}},
  \href{https://doi.org/10.1016/S0550-3213(02)01034-9}{\emph{Nucl. Phys.}
  {\bfseries B650} (2003) 229}
  [\href{https://arxiv.org/abs/hep-ph/0209154}{{\ttfamily hep-ph/0209154}}].

\bibitem{Banfi:2004yd}
A.~Banfi, G.~P. Salam and G.~Zanderighi, \emph{{Principles of general
  final-state resummation and automated implementation}},
  \href{https://doi.org/10.1088/1126-6708/2005/03/073}{\emph{JHEP} {\bfseries
  03} (2005) 073} [\href{https://arxiv.org/abs/hep-ph/0407286}{{\ttfamily
  hep-ph/0407286}}].

\bibitem{Becher:2006nr}
T.~Becher and M.~Neubert, \emph{{Threshold resummation in momentum space from
  effective field theory}},
  \href{https://doi.org/10.1103/PhysRevLett.97.082001}{\emph{Phys. Rev. Lett.}
  {\bfseries 97} (2006) 082001}
  [\href{https://arxiv.org/abs/hep-ph/0605050}{{\ttfamily hep-ph/0605050}}].

\bibitem{Luisoni:2015xha}
G.~Luisoni and S.~Marzani, \emph{{QCD resummation for hadronic final states}},
  \href{https://doi.org/10.1088/0954-3899/42/10/103101}{\emph{J. Phys.}
  {\bfseries G42} (2015) 103101}
  [\href{https://arxiv.org/abs/1505.04084}{{\ttfamily 1505.04084}}].

\bibitem{Bonocore:2014wua}
D.~Bonocore, E.~Laenen, L.~Magnea, L.~Vernazza and C.~D. White, \emph{{The
  method of regions and next-to-soft corrections in Drell–Yan production}},
  \href{https://doi.org/10.1016/j.physletb.2015.02.008}{\emph{Phys. Lett.}
  {\bfseries B742} (2015) 375}
  [\href{https://arxiv.org/abs/1410.6406}{{\ttfamily 1410.6406}}].

\bibitem{Bonocore:2015esa}
D.~Bonocore, E.~Laenen, L.~Magnea, S.~Melville, L.~Vernazza and C.~D. White,
  \emph{{A factorization approach to next-to-leading-power threshold
  logarithms}}, \href{https://doi.org/10.1007/JHEP06(2015)008}{\emph{JHEP}
  {\bfseries 06} (2015) 008}
  [\href{https://arxiv.org/abs/1503.05156}{{\ttfamily 1503.05156}}].

\bibitem{Bonocore:2016awd}
D.~Bonocore, E.~Laenen, L.~Magnea, L.~Vernazza and C.~D. White,
  \emph{{Non-abelian factorisation for next-to-leading-power threshold
  logarithms}}, \href{https://doi.org/10.1007/JHEP12(2016)121}{\emph{JHEP}
  {\bfseries 12} (2016) 121}
  [\href{https://arxiv.org/abs/1610.06842}{{\ttfamily 1610.06842}}].

\bibitem{DelDuca:2017twk}
V.~Del~Duca, E.~Laenen, L.~Magnea, L.~Vernazza and C.~D. White,
  \emph{{Universality of next-to-leading power threshold effects for colourless
  final states in hadronic collisions}},
  \href{https://doi.org/10.1007/JHEP11(2017)057}{\emph{JHEP} {\bfseries 11}
  (2017) 057} [\href{https://arxiv.org/abs/1706.04018}{{\ttfamily
  1706.04018}}].

\bibitem{Bahjat-Abbas:2018hpv}
N.~Bahjat-Abbas, J.~Sinninghe~Damsté, L.~Vernazza and C.~D. White, \emph{{On
  next-to-leading power threshold corrections in Drell-Yan production at
  N$^3$LO}}, \href{https://doi.org/10.1007/JHEP10(2018)144}{\emph{JHEP}
  {\bfseries 10} (2018) 144}
  [\href{https://arxiv.org/abs/1807.09246}{{\ttfamily 1807.09246}}].

\bibitem{Bhattacharya:2018vph}
A.~Bhattacharya, I.~Moult, I.~W. Stewart and G.~Vita, \emph{{Helicity Methods
  for High Multiplicity Subleading Soft and Collinear Limits}},
  \href{https://doi.org/10.1007/JHEP05(2019)192}{\emph{JHEP} {\bfseries 05}
  (2019) 192} [\href{https://arxiv.org/abs/1812.06950}{{\ttfamily
  1812.06950}}].

\bibitem{vanBeekveld:2019prq}
M.~van Beekveld, W.~Beenakker, E.~Laenen and C.~D. White,
  \emph{{Next-to-leading power threshold effects for inclusive and exclusive
  processes with final state jets}},
  \href{https://arxiv.org/abs/1905.08741}{{\ttfamily 1905.08741}}.

\bibitem{vanBeekveld:2019cks}
M.~van Beekveld, W.~Beenakker, R.~Basu, E.~Laenen, A.~Misra and P.~Motylinski,
  \emph{{Next-to-leading power threshold effects for resummed prompt photon
  production}},  \href{https://arxiv.org/abs/1905.11771}{{\ttfamily
  1905.11771}}.

\bibitem{Boughezal:2019ggi}
R.~Boughezal, A.~Isgrò and F.~Petriello, \emph{{Next-to-leading power
  corrections to $V+1$ jet production in $N$-jettiness subtraction}},
  \href{https://arxiv.org/abs/1907.12213}{{\ttfamily 1907.12213}}.

\bibitem{Moult:2016fqy}
I.~Moult, L.~Rothen, I.~W. Stewart, F.~J. Tackmann and H.~X. Zhu,
  \emph{{Subleading Power Corrections for N-Jettiness Subtractions}},
  \href{https://doi.org/10.1103/PhysRevD.95.074023}{\emph{Phys. Rev.}
  {\bfseries D95} (2017) 074023}
  [\href{https://arxiv.org/abs/1612.00450}{{\ttfamily 1612.00450}}].

\bibitem{Boughezal:2016zws}
R.~Boughezal, X.~Liu and F.~Petriello, \emph{{Power Corrections in the
  N-jettiness Subtraction Scheme}},
  \href{https://doi.org/10.1007/JHEP03(2017)160}{\emph{JHEP} {\bfseries 03}
  (2017) 160} [\href{https://arxiv.org/abs/1612.02911}{{\ttfamily
  1612.02911}}].

\bibitem{Moult:2017jsg}
I.~Moult, L.~Rothen, I.~W. Stewart, F.~J. Tackmann and H.~X. Zhu, \emph{{N
  -jettiness subtractions for $gg\to H$ at subleading power}},
  \href{https://doi.org/10.1103/PhysRevD.97.014013}{\emph{Phys. Rev.}
  {\bfseries D97} (2018) 014013}
  [\href{https://arxiv.org/abs/1710.03227}{{\ttfamily 1710.03227}}].

\bibitem{Boughezal:2018mvf}
R.~Boughezal, A.~Isgrò and F.~Petriello, \emph{{Next-to-leading-logarithmic
  power corrections for $N$-jettiness subtraction in color-singlet
  production}}, \href{https://doi.org/10.1103/PhysRevD.97.076006}{\emph{Phys.
  Rev.} {\bfseries D97} (2018) 076006}
  [\href{https://arxiv.org/abs/1802.00456}{{\ttfamily 1802.00456}}].

\bibitem{Ebert:2018lzn}
M.~A. Ebert, I.~Moult, I.~W. Stewart, F.~J. Tackmann, G.~Vita and H.~X. Zhu,
  \emph{{Power Corrections for N-Jettiness Subtractions at ${\cal
  O}(\alpha_s)$}}, \href{https://doi.org/10.1007/JHEP12(2018)084}{\emph{JHEP}
  {\bfseries 12} (2018) 084}
  [\href{https://arxiv.org/abs/1807.10764}{{\ttfamily 1807.10764}}].

\bibitem{Balitsky:2017flc}
I.~Balitsky and A.~Tarasov, \emph{{Higher-twist corrections to gluon TMD
  factorization}}, \href{https://doi.org/10.1007/JHEP07(2017)095}{\emph{JHEP}
  {\bfseries 07} (2017) 095}
  [\href{https://arxiv.org/abs/1706.01415}{{\ttfamily 1706.01415}}].

\bibitem{Balitsky:2017gis}
I.~Balitsky and A.~Tarasov, \emph{{Power corrections to TMD factorization for
  Z-boson production}},
  \href{https://doi.org/10.1007/JHEP05(2018)150}{\emph{JHEP} {\bfseries 05}
  (2018) 150} [\href{https://arxiv.org/abs/1712.09389}{{\ttfamily
  1712.09389}}].

\bibitem{Ebert:2018gsn}
M.~A. Ebert, I.~Moult, I.~W. Stewart, F.~J. Tackmann, G.~Vita and H.~X. Zhu,
  \emph{{Subleading power rapidity divergences and power corrections for
  q$_{T}$}}, \href{https://doi.org/10.1007/JHEP04(2019)123}{\emph{JHEP}
  {\bfseries 04} (2019) 123}
  [\href{https://arxiv.org/abs/1812.08189}{{\ttfamily 1812.08189}}].

\bibitem{Cieri:2019tfv}
L.~Cieri, C.~Oleari and M.~Rocco, \emph{{Higher-order power corrections in a
  transverse-momentum cut for colour-singlet production at NLO}},
  \href{https://doi.org/10.1140/epjc/s10052-019-7361-8}{\emph{Eur. Phys. J.}
  {\bfseries C79} (2019) 852}
  [\href{https://arxiv.org/abs/1906.09044}{{\ttfamily 1906.09044}}].

\bibitem{Hill:2004if}
R.~J. Hill, T.~Becher, S.~J. Lee and M.~Neubert, \emph{{Sudakov resummation for
  subleading SCET currents and heavy-to-light form-factors}},
  \href{https://doi.org/10.1088/1126-6708/2004/07/081}{\emph{JHEP} {\bfseries
  07} (2004) 081} [\href{https://arxiv.org/abs/hep-ph/0404217}{{\ttfamily
  hep-ph/0404217}}].

\bibitem{Beneke:2005gs}
M.~Beneke and D.~Yang, \emph{{Heavy-to-light B meson form-factors at large
  recoil energy: Spectator-scattering corrections}},
  \href{https://doi.org/10.1016/j.nuclphysb.2005.11.027}{\emph{Nucl. Phys.}
  {\bfseries B736} (2006) 34}
  [\href{https://arxiv.org/abs/hep-ph/0508250}{{\ttfamily hep-ph/0508250}}].

\bibitem{Freedman:2014uta}
S.~M. Freedman and R.~Goerke, \emph{{Renormalization of Subleading Dijet
  Operators in Soft-Collinear Effective Theory}},
  \href{https://doi.org/10.1103/PhysRevD.90.114010}{\emph{Phys. Rev.}
  {\bfseries D90} (2014) 114010}
  [\href{https://arxiv.org/abs/1408.6240}{{\ttfamily 1408.6240}}].

\bibitem{Goerke:2017lei}
R.~Goerke and M.~Inglis-Whalen, \emph{{Renormalization of dijet operators at
  order 1/Q$^{2}$ in soft-collinear effective theory}},
  \href{https://doi.org/10.1007/JHEP05(2018)023}{\emph{JHEP} {\bfseries 05}
  (2018) 023} [\href{https://arxiv.org/abs/1711.09147}{{\ttfamily
  1711.09147}}].

\bibitem{Beneke:2017ztn}
M.~Beneke, M.~Garny, R.~Szafron and J.~Wang, \emph{{Anomalous dimension of
  subleading-power N-jet operators}},
  \href{https://doi.org/10.1007/JHEP03(2018)001}{\emph{JHEP} {\bfseries 03}
  (2018) 001} [\href{https://arxiv.org/abs/1712.04416}{{\ttfamily
  1712.04416}}].

\bibitem{Beneke:2018rbh}
M.~Beneke, M.~Garny, R.~Szafron and J.~Wang, \emph{{Anomalous dimension of
  subleading-power $N$-jet operators. Part II}},
  \href{https://doi.org/10.1007/JHEP11(2018)112}{\emph{JHEP} {\bfseries 11}
  (2018) 112} [\href{https://arxiv.org/abs/1808.04742}{{\ttfamily
  1808.04742}}].

\bibitem{Beneke:2019kgv}
M.~Beneke, M.~Garny, R.~Szafron and J.~Wang, \emph{{Violation of the
  Kluberg-Stern-Zuber theorem in SCET}},
  \href{https://doi.org/10.1007/JHEP09(2019)101}{\emph{JHEP} {\bfseries 09}
  (2019) 101} [\href{https://arxiv.org/abs/1907.05463}{{\ttfamily
  1907.05463}}].

\bibitem{Moult:2018jjd}
I.~Moult, I.~W. Stewart, G.~Vita and H.~X. Zhu, \emph{{First Subleading Power
  Resummation for Event Shapes}},
  \href{https://doi.org/10.1007/JHEP08(2018)013}{\emph{JHEP} {\bfseries 08}
  (2018) 013} [\href{https://arxiv.org/abs/1804.04665}{{\ttfamily
  1804.04665}}].

\bibitem{Moult:2019mog}
I.~Moult, I.~W. Stewart and G.~Vita, \emph{{Subleading Power Factorization with
  Radiative Functions}},
  \href{https://doi.org/10.1007/JHEP11(2019)153}{\emph{JHEP} {\bfseries 11}
  (2019) 153} [\href{https://arxiv.org/abs/1905.07411}{{\ttfamily
  1905.07411}}].

\bibitem{Moult:2019uhz}
I.~Moult, I.~W. Stewart, G.~Vita and H.~X. Zhu, \emph{{The Soft Quark
  Sudakov}},  \href{https://arxiv.org/abs/1910.14038}{{\ttfamily 1910.14038}}.

\bibitem{Beneke:2018gvs}
M.~Beneke, A.~Broggio, M.~Garny, S.~Jaskiewicz, R.~Szafron, L.~Vernazza et~al.,
  \emph{{Leading-logarithmic threshold resummation of the Drell-Yan process at
  next-to-leading power}},
  \href{https://doi.org/10.1007/JHEP03(2019)043}{\emph{JHEP} {\bfseries 03}
  (2019) 043} [\href{https://arxiv.org/abs/1809.10631}{{\ttfamily
  1809.10631}}].

\bibitem{Bahjat-Abbas:2019fqa}
N.~Bahjat-Abbas, D.~Bonocore, J.~Sinninghe~Damsté, E.~Laenen, L.~Magnea,
  L.~Vernazza et~al., \emph{{Diagrammatic resummation of leading-logarithmic
  threshold effects at next-to-leading power}},
  \href{https://arxiv.org/abs/1905.13710}{{\ttfamily 1905.13710}}.

\bibitem{Beneke:2019mua}
M.~Beneke, M.~Garny, S.~Jaskiewicz, R.~Szafron, L.~Vernazza and J.~Wang,
  \emph{{Leading-logarithmic threshold resummation of Higgs production in gluon
  fusion at next-to-leading power}},
  \href{https://arxiv.org/abs/1910.12685}{{\ttfamily 1910.12685}}.

\bibitem{Beneke:2019oqx}
M.~Beneke, A.~Broggio, S.~Jaskiewicz and L.~Vernazza, \emph{{Threshold
  factorization of the Drell-Yan process at next-to-leading power}},
  \href{https://arxiv.org/abs/1912.01585}{{\ttfamily 1912.01585}}.

\bibitem{Moult:2019vou}
I.~Moult, G.~Vita and K.~Yan, \emph{{Subleading Power Resummation of Rapidity
  Logarithms: The Energy-Energy Correlator in $\mathcal{N}=4$ SYM}},
  \href{https://arxiv.org/abs/1912.02188}{{\ttfamily 1912.02188}}.

\bibitem{Jikia:1996bi}
G.~Jikia and A.~Tkabladze, \emph{{QCD corrections to heavy quark pair
  production in polarized gamma gamma collisions and the intermediate mass
  Higgs signal}}, \href{https://doi.org/10.1103/PhysRevD.54.2030}{\emph{Phys.
  Rev.} {\bfseries D54} (1996) 2030}
  [\href{https://arxiv.org/abs/hep-ph/9601384}{{\ttfamily hep-ph/9601384}}].

\bibitem{Fadin:1997sn}
V.~S. Fadin, V.~A. Khoze and A.~D. Martin, \emph{{Higgs studies in polarized
  gamma gamma collisions}},
  \href{https://doi.org/10.1103/PhysRevD.56.484}{\emph{Phys. Rev.} {\bfseries
  D56} (1997) 484} [\href{https://arxiv.org/abs/hep-ph/9703402}{{\ttfamily
  hep-ph/9703402}}].

\bibitem{Kotsky:1997rq}
M.~I. Kotsky and O.~I. Yakovlev, \emph{{On the resummation of double logarithms
  in the process Higgs $\to$ gamma gamma}},
  \href{https://doi.org/10.1016/S0370-2693(97)01260-4}{\emph{Phys. Lett.}
  {\bfseries B418} (1998) 335}
  [\href{https://arxiv.org/abs/hep-ph/9708485}{{\ttfamily hep-ph/9708485}}].

\bibitem{Akhoury:2001mz}
R.~Akhoury, H.~Wang and O.~I. Yakovlev, \emph{{On the Resummation of large QCD
  logarithms in Higgs $\to$ gamma gamma decay}},
  \href{https://doi.org/10.1103/PhysRevD.64.113008}{\emph{Phys. Rev.}
  {\bfseries D64} (2001) 113008}
  [\href{https://arxiv.org/abs/hep-ph/0102105}{{\ttfamily hep-ph/0102105}}].

\bibitem{Melnikov:2016emg}
K.~Melnikov and A.~Penin, \emph{{On the light quark mass effects in Higgs boson
  production in gluon fusion}},
  \href{https://doi.org/10.1007/JHEP05(2016)172}{\emph{JHEP} {\bfseries 05}
  (2016) 172} [\href{https://arxiv.org/abs/1602.09020}{{\ttfamily
  1602.09020}}].

\bibitem{Braaten:2017lxx}
E.~Braaten, H.~Zhang and J.-W. Zhang, \emph{{Mass Dependence of Higgs
  Production at Large Transverse Momentum}},
  \href{https://doi.org/10.1007/JHEP11(2017)127}{\emph{JHEP} {\bfseries 11}
  (2017) 127} [\href{https://arxiv.org/abs/1704.06620}{{\ttfamily
  1704.06620}}].

\bibitem{Braaten:2017ukc}
E.~Braaten, H.~Zhang and J.-W. Zhang, \emph{{Mass dependence of Higgs boson
  production at large transverse momentum through a bottom-quark loop}},
  \href{https://doi.org/10.1103/PhysRevD.97.096014}{\emph{Phys. Rev.}
  {\bfseries D97} (2018) 096014}
  [\href{https://arxiv.org/abs/1707.09857}{{\ttfamily 1707.09857}}].

\bibitem{Penin:2014msa}
A.~A. Penin, \emph{{High-Energy Limit of Quantum Electrodynamics beyond Sudakov
  Approximation}}, \href{https://doi.org/10.1016/j.physletb.2015.04.036,
  10.1016/j.physletb.2017.05.069, 10.1016/j.physletb.2015.10.035}{\emph{Phys.
  Lett.} {\bfseries B745} (2015) 69}
  [\href{https://arxiv.org/abs/1412.0671}{{\ttfamily 1412.0671}}].

\bibitem{Penin:2016wiw}
A.~A. Penin and N.~Zerf, \emph{{Two-loop Bhabha Scattering at High Energy
  beyond Leading Power Approximation}},
  \href{https://doi.org/10.1016/j.physletb.2017.05.073,
  10.1016/j.physletb.2016.07.077}{\emph{Phys. Lett.} {\bfseries B760} (2016)
  816} [\href{https://arxiv.org/abs/1606.06344}{{\ttfamily 1606.06344}}].

\bibitem{Liu:2017vkm}
T.~Liu and A.~A. Penin, \emph{{High-Energy Limit of QCD beyond the Sudakov
  Approximation}},
  \href{https://doi.org/10.1103/PhysRevLett.119.262001}{\emph{Phys. Rev. Lett.}
  {\bfseries 119} (2017) 262001}
  [\href{https://arxiv.org/abs/1709.01092}{{\ttfamily 1709.01092}}].

\bibitem{Alte:2018nbn}
S.~Alte, M.~König and M.~Neubert, \emph{{Effective Field Theory after a
  New-Physics Discovery}},
  \href{https://doi.org/10.1007/JHEP08(2018)095}{\emph{JHEP} {\bfseries 08}
  (2018) 095} [\href{https://arxiv.org/abs/1806.01278}{{\ttfamily
  1806.01278}}].

\bibitem{Liu:2018czl}
T.~Liu and A.~Penin, \emph{{High-Energy Limit of Mass-Suppressed Amplitudes in
  Gauge Theories}}, \href{https://doi.org/10.1007/JHEP11(2018)158}{\emph{JHEP}
  {\bfseries 11} (2018) 158}
  [\href{https://arxiv.org/abs/1809.04950}{{\ttfamily 1809.04950}}].

\bibitem{Alte:2019iug}
S.~Alte, M.~König and M.~Neubert, \emph{{Effective Theory for a Heavy Scalar
  Coupled to the SM via Vector-Like Quarks}},
  \href{https://doi.org/10.1140/epjc/s10052-019-6867-4}{\emph{Eur. Phys. J.}
  {\bfseries C79} (2019) 352}
  [\href{https://arxiv.org/abs/1902.04593}{{\ttfamily 1902.04593}}].

\bibitem{Inoue:1994jq}
M.~Inoue, R.~Najima, T.~Oka and J.~Saito, \emph{{QCD corrections to two photon
  decay of the Higgs boson and its reverse process}},
  \href{https://doi.org/10.1142/S0217732394001003}{\emph{Mod. Phys. Lett.}
  {\bfseries A9} (1994) 1189}.

\bibitem{Spira:1995rr}
M.~Spira, A.~Djouadi, D.~Graudenz and P.~M. Zerwas, \emph{{Higgs boson
  production at the LHC}},
  \href{https://doi.org/10.1016/0550-3213(95)00379-7}{\emph{Nucl. Phys.}
  {\bfseries B453} (1995) 17}
  [\href{https://arxiv.org/abs/hep-ph/9504378}{{\ttfamily hep-ph/9504378}}].

\bibitem{Fleischer:2004vb}
J.~Fleischer, O.~V. Tarasov and V.~O. Tarasov, \emph{{Analytical result for the
  two loop QCD correction to the decay H $\to$ 2 gamma}},
  \href{https://doi.org/10.1016/j.physletb.2004.01.063}{\emph{Phys. Lett.}
  {\bfseries B584} (2004) 294}
  [\href{https://arxiv.org/abs/hep-ph/0401090}{{\ttfamily hep-ph/0401090}}].

\bibitem{Harlander:2005rq}
R.~Harlander and P.~Kant, \emph{{Higgs production and decay: Analytic results
  at next-to-leading order QCD}},
  \href{https://doi.org/10.1088/1126-6708/2005/12/015}{\emph{JHEP} {\bfseries
  12} (2005) 015} [\href{https://arxiv.org/abs/hep-ph/0509189}{{\ttfamily
  hep-ph/0509189}}].

\bibitem{Anastasiou:2006hc}
C.~Anastasiou, S.~Beerli, S.~Bucherer, A.~Daleo and Z.~Kunszt, \emph{{Two-loop
  amplitudes and master integrals for the production of a Higgs boson via a
  massive quark and a scalar-quark loop}},
  \href{https://doi.org/10.1088/1126-6708/2007/01/082}{\emph{JHEP} {\bfseries
  01} (2007) 082} [\href{https://arxiv.org/abs/hep-ph/0611236}{{\ttfamily
  hep-ph/0611236}}].

\bibitem{Aglietti:2006tp}
U.~Aglietti, R.~Bonciani, G.~Degrassi and A.~Vicini, \emph{{Analytic Results
  for Virtual QCD Corrections to Higgs Production and Decay}},
  \href{https://doi.org/10.1088/1126-6708/2007/01/021}{\emph{JHEP} {\bfseries
  01} (2007) 021} [\href{https://arxiv.org/abs/hep-ph/0611266}{{\ttfamily
  hep-ph/0611266}}].

\bibitem{Spira:2016ztx}
M.~Spira, \emph{{Higgs Boson Production and Decay at Hadron Colliders}},
  \href{https://doi.org/10.1016/j.ppnp.2017.04.001}{\emph{Prog. Part. Nucl.
  Phys.} {\bfseries 95} (2017) 98}
  [\href{https://arxiv.org/abs/1612.07651}{{\ttfamily 1612.07651}}].

\bibitem{Liu:2019oav}
Z.~L. Liu and M.~Neubert, \emph{{Factorization at Subleading Power and
  Endpoint-Divergent Convolutions in $h\to\gamma\gamma$ Decay}},
  \href{https://arxiv.org/abs/1912.08818}{{\ttfamily 1912.08818}}.

\bibitem{Smirnov:1997gx}
V.~A. Smirnov, \emph{{Asymptotic expansions of two loop Feynman diagrams in the
  Sudakov limit}},
  \href{https://doi.org/10.1016/S0370-2693(97)00545-5}{\emph{Phys. Lett.}
  {\bfseries B404} (1997) 101}
  [\href{https://arxiv.org/abs/hep-ph/9703357}{{\ttfamily hep-ph/9703357}}].

\bibitem{Beneke:1997zp}
M.~Beneke and V.~A. Smirnov, \emph{{Asymptotic expansion of Feynman integrals
  near threshold}},
  \href{https://doi.org/10.1016/S0550-3213(98)00138-2}{\emph{Nucl. Phys.}
  {\bfseries B522} (1998) 321}
  [\href{https://arxiv.org/abs/hep-ph/9711391}{{\ttfamily hep-ph/9711391}}].

\bibitem{Bauer:2000ew}
C.~W. Bauer, S.~Fleming and M.~E. Luke, \emph{{Summing Sudakov logarithms in B
  $\to$ $X_s$ + gamma in effective field theory}},
  \href{https://doi.org/10.1103/PhysRevD.63.014006}{\emph{Phys. Rev.}
  {\bfseries D63} (2000) 014006}
  [\href{https://arxiv.org/abs/hep-ph/0005275}{{\ttfamily hep-ph/0005275}}].

\bibitem{Bauer:2000yr}
C.~W. Bauer, S.~Fleming, D.~Pirjol and I.~W. Stewart, \emph{{An Effective field
  theory for collinear and soft gluons: Heavy to light decays}},
  \href{https://doi.org/10.1103/PhysRevD.63.114020}{\emph{Phys.Rev.} {\bfseries
  D63} (2001) 114020} [\href{https://arxiv.org/abs/hep-ph/0011336}{{\ttfamily
  hep-ph/0011336}}].

\bibitem{Bauer:2001yt}
C.~W. Bauer, D.~Pirjol and I.~W. Stewart, \emph{{Soft collinear factorization
  in effective field theory}},
  \href{https://doi.org/10.1103/PhysRevD.65.054022}{\emph{Phys.Rev.} {\bfseries
  D65} (2002) 054022} [\href{https://arxiv.org/abs/hep-ph/0109045}{{\ttfamily
  hep-ph/0109045}}].

\bibitem{Bauer:2002nz}
C.~W. Bauer, S.~Fleming, D.~Pirjol, I.~Z. Rothstein and I.~W. Stewart,
  \emph{{Hard scattering factorization from effective field theory}},
  \href{https://doi.org/10.1103/PhysRevD.66.014017}{\emph{Phys. Rev.}
  {\bfseries D66} (2002) 014017}
  [\href{https://arxiv.org/abs/hep-ph/0202088}{{\ttfamily hep-ph/0202088}}].

\bibitem{Beneke:2002ph}
M.~Beneke, A.~Chapovsky, M.~Diehl and T.~Feldmann, \emph{{Soft collinear
  effective theory and heavy to light currents beyond leading power}},
  \href{https://doi.org/10.1016/S0550-3213(02)00687-9}{\emph{Nucl.Phys.}
  {\bfseries B643} (2002) 431}
  [\href{https://arxiv.org/abs/hep-ph/0206152}{{\ttfamily hep-ph/0206152}}].

\bibitem{Leibovich:2003jd}
A.~K. Leibovich, Z.~Ligeti and M.~B. Wise, \emph{{Comment on quark masses in
  SCET}}, \href{https://doi.org/10.1016/S0370-2693(03)00565-3}{\emph{Phys.
  Lett.} {\bfseries B564} (2003) 231}
  [\href{https://arxiv.org/abs/hep-ph/0303099}{{\ttfamily hep-ph/0303099}}].

\bibitem{Chay:2004zn}
J.~Chay, C.~Kim, Y.~G. Kim and J.-P. Lee, \emph{{Soft Wilson lines in
  soft-collinear effective theory}},
  \href{https://doi.org/10.1103/PhysRevD.71.056001}{\emph{Phys. Rev.}
  {\bfseries D71} (2005) 056001}
  [\href{https://arxiv.org/abs/hep-ph/0412110}{{\ttfamily hep-ph/0412110}}].

\bibitem{Manohar:2006nz}
A.~V. Manohar and I.~W. Stewart, \emph{{The Zero-Bin and Mode Factorization in
  Quantum Field Theory}},
  \href{https://doi.org/10.1103/PhysRevD.76.074002}{\emph{Phys. Rev.}
  {\bfseries D76} (2007) 074002}
  [\href{https://arxiv.org/abs/hep-ph/0605001}{{\ttfamily hep-ph/0605001}}].

\bibitem{Idilbi:2007ff}
A.~Idilbi and T.~Mehen, \emph{{On the equivalence of soft and zero-bin
  subtractions}}, \href{https://doi.org/10.1103/PhysRevD.75.114017}{\emph{Phys.
  Rev.} {\bfseries D75} (2007) 114017}
  [\href{https://arxiv.org/abs/hep-ph/0702022}{{\ttfamily hep-ph/0702022}}].

\bibitem{Idilbi:2007yi}
A.~Idilbi and T.~Mehen, \emph{{Demonstration of the equivalence of soft and
  zero-bin subtractions}},
  \href{https://doi.org/10.1103/PhysRevD.76.094015}{\emph{Phys. Rev.}
  {\bfseries D76} (2007) 094015}
  [\href{https://arxiv.org/abs/0707.1101}{{\ttfamily 0707.1101}}].

\bibitem{Chiu:2009yx}
J.-y. Chiu, A.~Fuhrer, A.~H. Hoang, R.~Kelley and A.~V. Manohar,
  \emph{{Soft-Collinear Factorization and Zero-Bin Subtractions}},
  \href{https://doi.org/10.1103/PhysRevD.79.053007}{\emph{Phys. Rev.}
  {\bfseries D79} (2009) 053007}
  [\href{https://arxiv.org/abs/0901.1332}{{\ttfamily 0901.1332}}].

\bibitem{Becher:2012xx}
T.~Becher and G.~Bell, \emph{{Analytic regularization in Soft-Collinear
  Effective Theory}},
  \href{https://doi.org/10.1016/j.physletb.2012.05.016}{\emph{Phys. Lett.}
  {\bfseries B713} (2012) 41}
  [\href{https://arxiv.org/abs/1112.3907}{{\ttfamily 1112.3907}}].

\bibitem{Becher:2009qa}
T.~Becher and M.~Neubert, \emph{{On the Structure of Infrared Singularities of
  Gauge-Theory Amplitudes}},
  \href{https://doi.org/10.1088/1126-6708/2009/06/081,
  10.1007/JHEP11(2013)024}{\emph{JHEP} {\bfseries 06} (2009) 081}
  [\href{https://arxiv.org/abs/0903.1126}{{\ttfamily 0903.1126}}].

\bibitem{Neubert:2004dd}
M.~Neubert, \emph{{Renormalization-group improved calculation of the B $\to$
  $X_s$ gamma branching ratio}},
  \href{https://doi.org/10.1140/epjc/s2005-02141-1}{\emph{Eur. Phys. J.}
  {\bfseries C40} (2005) 165}
  [\href{https://arxiv.org/abs/hep-ph/0408179}{{\ttfamily hep-ph/0408179}}].

\end{thebibliography}\endgroup
\bibliographystyle{JHEP}


\end{document}